\documentclass[slac_one]{revtex4}
\usepackage{graphicx}
\usepackage{fancyhdr}
\begin{document}

\title{{\small{2005 ALCPG \& ILC Workshops - Snowmass,
U.S.A.}}\\ %% Please keep this conference title here
\vspace{12pt}
Understanding Light: Why we need a terascale photon collider}

\author{Zack Sullivan}
\affiliation{High Energy Physics Division, Argonne National Laboratory,
Argonne, IL 60439, USA}

\begin{abstract}
We do not understand light.  I argue that a terascale photon collider
is necessary to determine the structure of the photon at 100 GeV.
Uncertainties in photon parton distribution functions lead to cross
section predictions that vary by a factor of 5.  This limits our
ability to predict how well we can perform precision measurements, e.g.,
extracting the width of Higgs into two photons.  These uncertainties
will only be resolved by measuring the gluonic structure of the photon
\textit{in situ}.
\end{abstract}

\maketitle

\thispagestyle{fancy}

\section{INTRODUCTION\label{sec:intro}}

A compelling motivation for a construction of a photon collider is the
precision measurement of electroweak observables
\cite{Badelek:2001xb}.  In particular, two of the most important are
the measurement of the Higgs total width, and the width of Higgs into
two photons.  The later is directly proportional to the cross section
through two photons, and hence a measure of the cross section is a
measure of the partial width.  A complete study of the cross section
into two $b$ jets ($\gamma\gamma\to h, A\to bb$) is presented at this
workshop \cite{Niezurawski:2002aq}.  That analysis contains a full
NLO treatment of backgrounds, a realistic photon spectrum, and
detector simulation.  They estimate that the ``resolved'' hadronic
component of photons contributes about 15\% to the background to Higgs
production at a mass of 120 GeV.  In this analysis, I demonstrate that
the normalization and shape of this resolved-photon background to
Higgs production are only known to a factor of 3--4.  Hence, the
photons are interesting in themselves, and not just as a tool to probe
other physics.

The complete study of resolved photons near threshold is difficult to
model properly.  However, a simple leading order (LO) analysis of
$bb$-dijet production is enough to quantify the uncertainties
described in Sec.\ \ref{sec:uncertain} below.  We begin by plotting in
Fig.\ \ref{fig:totbb} the $bb$-dijet cross section as a function of
$bb$-invariant mass $M_{bb}$.  The events were generated using a flat
photon energy spectrum \cite{Niezurawski:2002aq} of 25--200 GeV/beam.
This is a good approximation to the full non-linear spectrum coming
from photons back-scattered off of 250 GeV electrons.  The two $b$
jets each have $E_{Tb}>40$ GeV, $|\eta_b|<4$, and a separation $\Delta
R>0.1$.  Looser cuts would slightly increase the $\gamma\gamma$
contribution near 80 GeV, but would greatly increase the $\gamma g$
contribution.

\begin{figure*}[tb]
\centering
\includegraphics[width=3.25in]{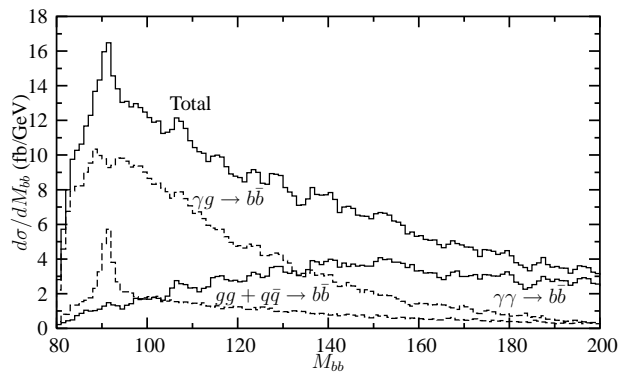}
\caption{Breakdown of contributions to the $bb$-dijet cross section
vs.\ dijet invariant mass $M_{bb}$.}\label{fig:totbb}
\end{figure*}

The lowest solid curve in Fig.\ \ref{fig:totbb} is the distribution
for the ``direct'' collision of two real photons $\gamma\gamma\to
b\bar b$.  Rising above the real photon contribution is a large peak
at the $Z$ mass coming from resolved-resolved quark annihilation into
a real $Z$ boson.  These quarks (and also gluons) come from hard
interactions that ``resolve'' the hadronic structure of the photon.
Both contributions are swamped by the single-resolved cross section
from photon-gluon fusion $\gamma g\to b\bar b$.  This is a generic
feature at any photon collider where the final state invariant mass is
less than $\sim W_{\textrm{max}}/3$.  Hence, a 120 GeV Higgs at a 500
GeV $ee$ collider, which produces a $\gamma\gamma$ collider with
$W_{\textrm{max}}\approx 400$ GeV, would have a huge background from
resolved photons.

The concern that resolved photons would be a large background to
Higgs production was pointed out in Ref.\ \cite{Eboli:1993jt}, but was
expected to be reduced by careful tuning of the electron beam energy
and creation of polarized photons.  More recent studies
\cite{Niezurawski:2002aq} have found a flatter spectrum of photon energies
than previously assumed, and NLO effects appear to reduce the usefulness of
polarized photon beams in reducing the background.

\section{LARGE UNCERTAINTIES IN THE $bb$-DIJET CROSS
SECTION\label{sec:uncertain}}

I turn to the estimation of the uncertainties for the resolved photon
cross sections due purely to our understanding of the photon parton
distribution functions (PDFs).  The $bb$ cross section is calculated using
the 8 auxiliary PDFs given by the CJK group, and the ``modified
tolerance method'' \cite{Sullivan:2002jt} with the default CJK2
tolerance of 10 \cite{Jankowski:2003mt,Cornet:2004ng}.

\begin{figure*}[tb]
\centering
\includegraphics[width=3.25in]{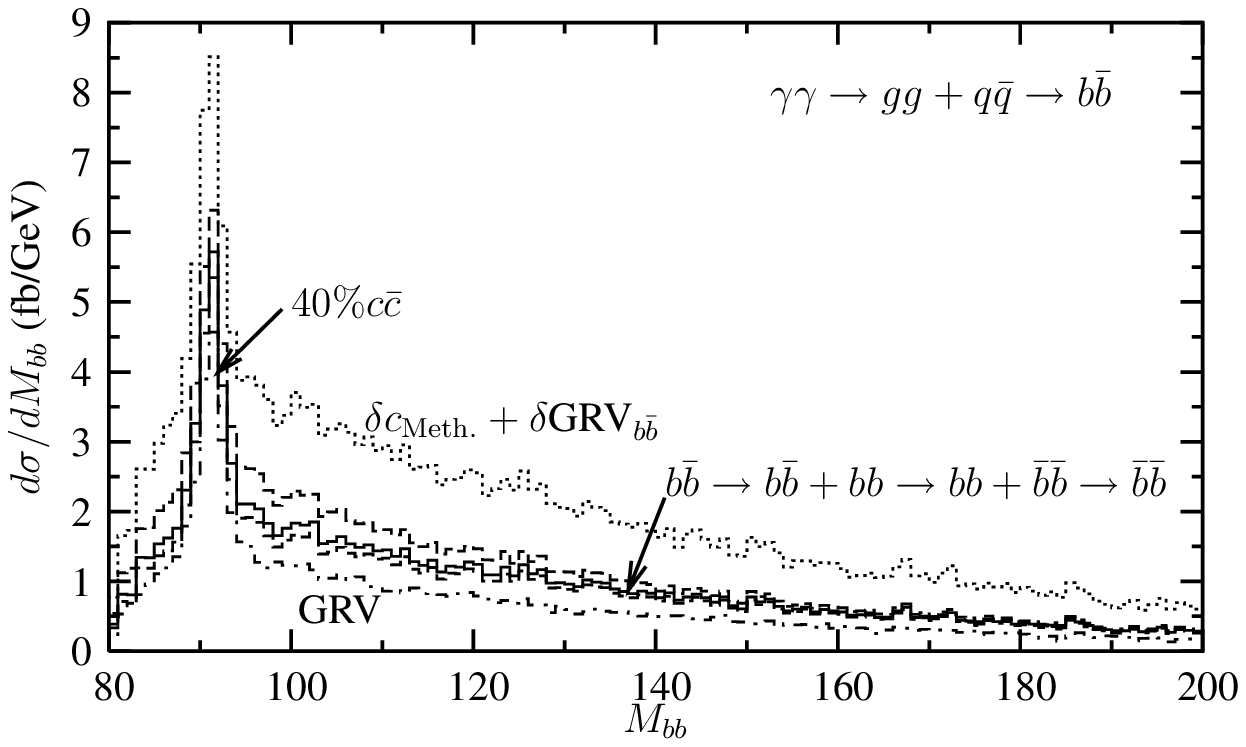}\hspace*{0.25in}\includegraphics[width=3.25in]{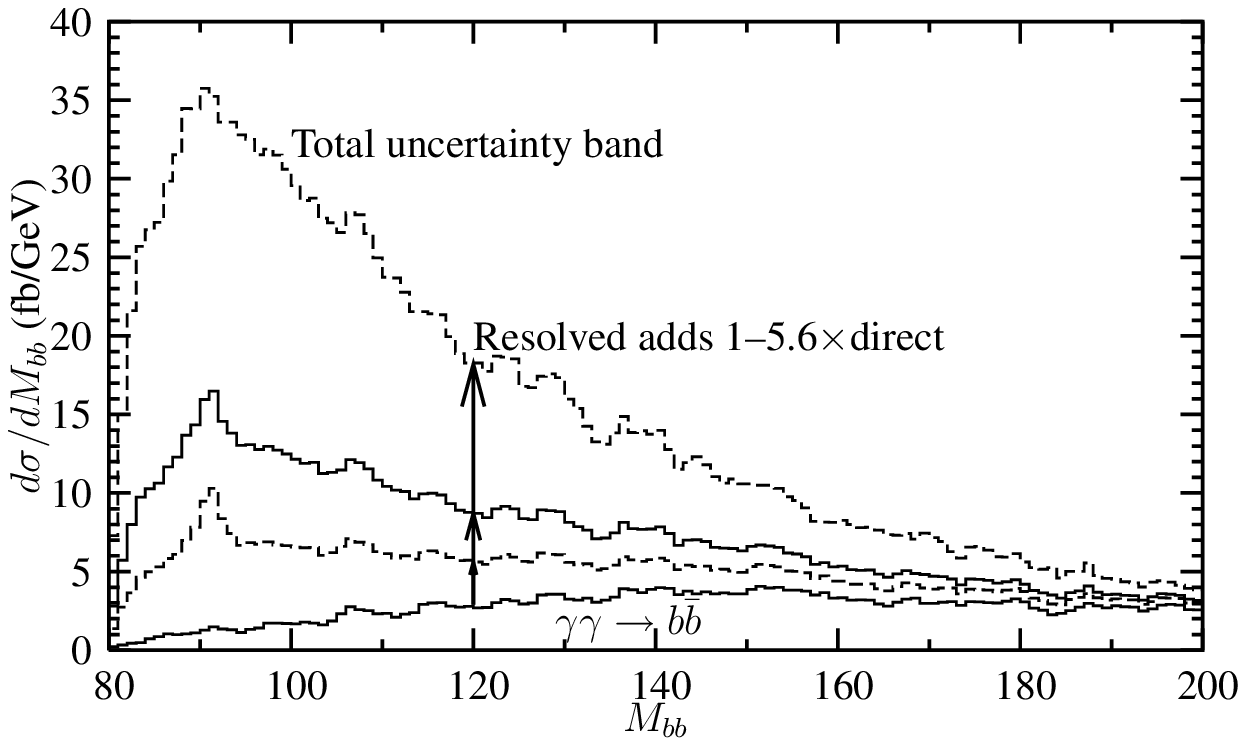}\\
{ \hspace*{0.25in}(a)\hspace*{3.35in}(b) }
\caption{(a) Uncertainty in the resolved-resolved contribution to
$bb$-dijet cross section vs.\ dijet invariant mass
$M_{bb}$.\\
(b) Range of theoretical predictions for $bb$-dijet cross section
vs.\ dijet invariant mass $M_{bb}$.}\label{fig:xxerr}
\end{figure*}

The results for the resolved-resolved contribution are shown in Fig.\
\ref{fig:xxerr}(a).  The solid line is the central value from Fig.\
\ref{fig:totbb}, and the dashed lines are uncertainty in the
prediction using the modified tolerance method.  Based on this
calculation alone, it would appear that the cross section is
well-constrained.  Also shown with dash-dots is the prediction of using
the GRV LO \cite{Gluck:1991jc} PDFs.  We notice that the GRV prediction
is far below the supposed lower limit of the uncertainty.  This indicates
that we should be careful in interpreting these uncertainties.

In order to get a better estimate of the uncertainty, I focus on two
features of Fig.\ \ref{fig:xxerr}(a). First, 40\% of the peak of the
distribution comes from $c\bar c$ annihilation through the $Z$ pole.
Second, the long tail to high invariant mass is mostly due to
rescattering of real $b$ quarks from the photon ($b\bar b\to b\bar b$,
$bb\to bb$, and $\bar b\bar b\to \bar b\bar b$).  Both the $c$ and $b$
PDFs are significantly larger at large $x$ than the GRV PDFs.
Furthermore, it was shown \cite{Jankowski:2003mt} that even at 4 GeV,
the tolerance method used to produce the PDFs begins to diverge from a
more accurate Lagrange multiplier result --- implying that the
some of the underlying assumptions used to calculate the PDF
uncertainties are questionable.  Therefore, a dotted curve also appears
on Fig.\ \ref{fig:xxerr}(a) that estimates a rough upper limit on the
error based on the difference between the GRV and CJK2 results.  This
corresponds to about a 50\% uncertainty near the $Z$ peak, and a
factor of 2 in the tail.

While the uncertainty in the resolved-resolved cross section is
interesting, the dominant cross section came from direct-resolved
collisions.  Unfortunately, the gluon is barely constrained.  Much of
this has to do with the need to subtract nonperturbative physics at
low energy scales in order to get at the perturbative partons.  One
result is that the tolerance parameter $T$ increases to over 100 for
the $g$ PDF near 100 GeV.  This leads to at least a factor of 3
uncertainty in the $\gamma g\to b\bar b$ cross section.

Putting this all together, we see in Fig.\ \ref{fig:xxerr}(b) that the
theoretical prediction for the $bb$-dijet cross section ranges from 5--20
fb/GeV at 120 GeV.  In this figure I have used the tolerance method
results, but a more conservative result would combine the differences
between PDF fits to predict something like 3--25 fb/GeV.  The
uncertainties themselves are only approximate, given the poor PDF fits
to the data.

The large uncertainty in the $bb$-dijet cross section should not be
surprising.  Measurements from experiments at both the HERA collider
at DESY \cite{Behnke:2004yk} and the LEP collider at CERN
\cite{Sefkow:2001bs} exhibit a factor of 2--3 excess of events over
the NLO theoretical predictions.  It seems likely that at least a part
of this excess may be attributed to a larger than expected gluon PDF.
This can easily be determined at a real photon collider with cuts like
those above, that give cross sections dominated by photon-gluon fusion.

\section{CONCLUSIONS\label{sec:conclusion}}

The difficulty in predicting the $bb$-dijet cross section presents an
exciting opportunity.  Given the dominance of the photon-gluon fusion
cross section, the first measurement of $b$ production can quickly pin
down the elusive gluonic component of the photon.  If the $Z$ peak can
be observed, the charm PDF can also be pinned down.  The $b$ PDF
might be measured from the resolved-resolved tail above the $Z$ peak.
One experimental challenge will be distinguishing charm quarks that
fake $b$ jets, but demanding two $b$ tags may suppress it enough to
pull out the physics.  Once these PDFs are measured, an accurate
measurement of the Higgs coupling to two photons can be made.

In this paper I have focused on a photon-photon collider, but another
option may be to use a photon-electron collider to measure the photon
structure.  It is more difficult to cleanly extract the gluon PDF in a
$\gamma$--$e$ collision than a $\gamma$--$\gamma$ collision, because
the cross section at high invariant mass ($>20$ GeV) is smaller, the
decay products tend to be boosted more forward into less-well
instrumented regions of the detector, and an additional deconvolution
must be performed to remove the effect of extracting an almost-real
photon from the electron.  Nevertheless, this option should be
examined in detail as it may be simpler to construct a $\gamma$--$e$
collider.

Today we do not understand light, but with a terascale photon
collider, we will.

\begin{acknowledgments}
This work was supported by an FY2005 grant from the Argonne Theory
Institute, and by the U.~S.~Department of Energy under Contract No.\
W-31-109-ENG-38.
\end{acknowledgments}

\end{document}